\documentclass[]{raa}
\usepackage{graphicx,times,amsmath}

\begin{document}

\title{Equilibria of a charged artificial satellite subject to gravitational
and Lorentz torques}
\author{Yehia A. Abdel-Aziz \inst{1},\inst{2} \and Muhammad Shoaib \inst{2} }
\date{Received~~2009 month day; accepted~~2009~~month day}

\volnopage{Vol.0 (200x) No.0, 000--000}
\setcounter{page}{1} 

\institute{ National Research Institute of Astronomy and Geophysics (UNRIG), Helena, Cairo, Egypt;
{\it yehia@nriag.sci.eg}\\
\and
University of Hail, Department of Mathematics, PO BOX
2440, Kingdom of Saudi Arabia, {\it safridi@gmail.com}}

\abstract{ Attitude Dynamics of a rigid artificial satellite subject to gravity gradient and Lorentz torques in a circular orbit is considered. Lorentz torque is developed on the basis of the electrodynamic effects of the Lorentz force acting on the charged satellite's surface. We assume that the satellite is moving  in  Low Earth Orbit (LEO)  in the geomagnetic field which is considered as a dipole model. Our model of the torque due to the Lorentz force is developed for a  general shape of artificial satellite, and the nonlinear differential equations of Euler are used to describe its attitude orientation. All equilibrium positions are determined and {their} existence conditions  are obtained. The numerical results show that the charge $q$ and radius $\rho_0$ of the charged center of satellite provide a  certain type of  semi passive control for the attitude  of  satellite. The technique for such kind of control would be  to increase or decrease the electrostatic radiation screening of the satellite. The results  {obtained} confirm that the change in charge can effect the magnitude of the Lorentz torque, which may affect the satellite's control. Moreover, the relation between the magnitude of the Lorentz torque and  inclination of the orbits is investigated.
} \authorrunning{Yehia A. Abdel-Aziz \& M. Shoaib }
\titlerunning{Equilibria of a charged artificial satellite subject to gravitational
and Lorentz torques } 

.

\maketitle

\section{Introduction}

Artificial satellite moving in Low Earth Orbit (LEO) or High Earth Orbit
(HEM) naturally tends to accumulate electrostatic charge.
{Ambient plasma and photoelectric
effect can produce Lorentz force in LEO. The spacecraft plasma interaction
is the main source for spacecraft charging. Due
to plasma interactions spacecraft surface charging  is the major source of spacecraft anomalies (Garget
1981, Garget et.all 1984).  In some cases the accumulation of  electrostatic charge  affect the instruments and other devices onboard the satellite,}
which may ultimately lead to difficulties in operating the satellite.
 {For example the newly launched LARES
satellite can be effected by electrostatic charging (Chinoline et.al, 2012).
Similarly, the Space Shuttle has been investigated for charging (Bile et.al,
1995).} Different research efforts have led to the development of technology
of active mitigation of the satellite charging through the control of
charge.  {The effect of electrostatic charge may negatively impact the error budget of satellites, designed for
experiments of fundamental physics, by damaging onboard electronic
instruments or by interfering with scientific measurement. The damage to
electronic instruments is rare but may be harmful in many ways. The
interference with scientific measurement is very common due to spacecraft
charging. See references Everts et.al. (2011), Worded and Everts (2013),
Nobile et al. (2009), Orion (2009) and Orion, et al. (2004) and the
references there in.}

 {Coprophilagy (\cite{VokRouhlicky1989}) and
Salad \& Ismaili (\cite{Saad2010}) determined the orbital effects of the
Lorentz force on the motion of an electrically charged artificial satellite
moving in the Earth's magnetic field. The influence of the geomagnetic field
manifests itself predominantly by Lorentz force. Then in 1990 Coprophilagy
studied variation in the orbital elements due to Lorentz force with
variation in natural charge. Pollack et.al (2010) show that Lorentz force
may be used to save substantial propellant in inclination change maneuvers.
Heelamon et.al (2012) show that the effect of electric dipole moment induced
by the high altitude Earth electric field is very small as compared to the
electromagnetic effect. Png and Gao (2012) show that Lorentz force can be
implemented for $J_2$ invariant formation given that the deputy spacecraft
has electrostatic charge.} Therefore Lorentz force is a possible means for
charging and thus controlling spacecraft orbits without consuming
propellant. Peck (\cite{peck2005}) was the first to introduce a control
scheme. The spacecraft orbits accelerated by the Lorentz force are termed
Lorentz --augmented orbits, because Lorentz force cannot complectly replace
the traditional rocket propulsion. After Peck (\cite{peck2005}) a series of
papers (King et.al \cite{king2003}; Ataraxy \& Schauder \cite{Natarjan2006};
Streetcar \& Peck \cite{streetman2007}; Utahn \& Hiroshima \cite{utako2008};
Hiroshima et.al \cite{hiroshi2009}) applied charge control techniques to the
utilization of Lorentz forces for satellite orbit control.

Abide-Ariz (\cite{yehia2007}) have studied the stability of equilibrium
position due to Lorentz torque in the case of uniform magnetic field and
cylindrical shape for an artificial satellite. Kawakawa et.al. (\cite%
{yamakawa2012}) investigated the attitude motion of a charged pendulum
satellite having the shape of a dumbbell pendulum due to Lorentz torque.
Their study of stability of equilibrium points is focused only on pitch
position within the equatorial plane.

In this paper, we are concerned with the attitude motion of an artificial
satellite of general shape moving in a circular orbit under gravity gradient
torque and Lorentz torque. Euler equations will be used to describe the
attitude dynamics of the satellite. Determination of equilibrium orientation
of a satellite under the action of gravitational and Lorentz torques is one
of the basic problems of this paper. Finally, we will analyze the
equilibrium positions based on control of the charged center of the
satellite relative to its center of mass and the amount of charge.

 {Before we move onto the next section to formulate the problem in question we would like to point out that electromagnetic effects caused by a Lorentz force on satellites moving in the gravitational field of the Earth, subject of this paper, are not to be confused with purely gravitational effects, dubbed "gravitomagentic" arising from general relativity. They are widely diffused in literature (Mashhoon et.al 2001, Mashhoon 2007, and Orion \& Lichtenegger 2005). The name "gravitomagnetic" is due to a purely formal resemblance of the Lense-Thirring effects, arising in stationary space-times generated by stationary mass-energy currents such as a rotating planet, with the linear equations of electromagnetism by Maxwell and with the Lorentz force acting on electrically charged bodies moving in a magnetic field (Orion et al 2011, Orion et al 2002, 	Renzetti 2013, Mashhoon 2013, and Lichtenegger et.al 2006).
}

\section{ Formulation of the Problem}

A rigid spacecraft is considered whose center of mass moves in the Newtonian
central gravitational field of the earth in a circular orbit of radius $r$.
We suppose that the spacecraft is equipped with an electrostatically charged
protective shield, having an intrinsic magnetic moment. The rotational
motion of the spacecraft about its center of mass will be analyzed,
considering the influence of gravity gradient torque $T_{G}$ and the torque $%
{\ T}_{L}$ due to Lorentz forces respectively. The torque ${T}_{L}$ results
from the interaction of the geomagnetic field with the charged screen of the
electrostatic shield.

The rotational motion of the satellite relative to its center of mass is
investigated in the orbital coordinate system $C_{{x_o}{y_o}{z_0}}$ with $%
C_{x_o}$ tangent to the orbit in the direction of motion, $C_{y_o}$ lies
along the normal to the orbital plane, and $C_{z_o}$ lies along the radius
vector $r$ of the point $O_E$ relative to the center of the Earth. The
investigation is carried out assuming the rotation of the orbital coordinate
system relative to the inertial system with the angular velocity $\Omega$.
As an inertial coordinate system, the system $O_{XYZ}$ is taken, whose axis $%
OZ(k)$ is directed along the axis of the Earth's rotation, the axis $OX(i)$
is directed toward the ascending node of the orbit, and the plane coincides
with the equatorial plane. Also, we assume that the satellite's principal
axes of inertia $C_{{x_b}{y_b}{z_b}}$ are rigidly fixed to a satellite $%
(i_b,j_b,k_b)$. The satellite's attitude may be described in several ways,
in this paper the attitude will be described by the angle of yaw $\psi \, $
the angle of pitch $\theta $ , and the angle of roll $\varphi $, between the
satellite's $C_{{x_b}{y_b}{z_b}}$ and the set of reference axes $O_{XYZ}$.
The three angles are obtained by rotating satellite axes from an attitude
coinciding with the reference axes to describe attitude in the following way:

- Allow a rotation $\psi \, $ about z-axis

- About the newly displaced y-axis, rotate through $\theta $

- Finally allow a rotation $\varphi $ about the final position of the x-axis

Although the angles $\psi $, $\theta $ and $\varphi $ are often referred to
as Euler angles, they differ from classical Euler angles in that only
rotation takes place about each axis, whereas in the classical Euler angular
coordinates, two rotations are made about the z-axis. The relation between
the orbital coordinate system and reference system $O_{XYZ}$ is determined
as below.

\begin{equation}
\begin{array}{l}
{\hat{\imath}=-\sin u{\vec{\alpha}+}\cos u{\vec{\gamma},}} \\
{{\hat{\jmath}}=\cos i\cos u{\vec{\alpha}-}\sin i{\vec{\beta}+}\cos i\sin u{%
\vec{\gamma},}} \\
{{\hat{\kappa}}=\sin i\cos u{\vec{\alpha}+}\cos i{\vec{\beta}+}\sin i\sin u{%
\vec{\gamma},}}%
\end{array}
\label{GrindEQ__1_}
\end{equation}%
where $i$ is the orbital inclination and  { $u=\Omega t+u_{0}$ is the argument
of latitude, $\Omega $ is the orbital angular velocity of the satellite's
center of mass, $u_{0}$ is the initial latitude and $\vec{\alpha}$,$\vec{%
\beta}$,$\vec{\gamma}$ are unit vectors along the axes of the orbital
coordinate system.} These vectors are the different directions of the tangent
to plane of the orbit, its radius and the normal of the orbit respectively
(Gerlach \cite{gerlach1965}).

The relationship between the reference frames $C_{{x_{b}}{y_{b}}{z_{b}}}$
and $C_{{x_{o}}{y_{o}}{z_{0}}}$ is given by the matrix $A$ which is the
matrix of  {unitary vectors} $\alpha
_{i},\,\,\beta _{i},\,\,\gamma _{i}$,$({i}=1,2,3).$

\begin{equation}
A=%
\begin{pmatrix}
\alpha _{1} & \alpha _{2} & \alpha _{3} \\
\beta _{1} & \beta _{2} & \beta _{3} \\
\gamma _{1} & \gamma _{2} & \gamma _{3}%
\end{pmatrix}%
,
\end{equation}%
where

\begin{equation}
\begin{array}{l}
{\alpha _{1}=\cos \theta \cos \psi ,} \\
{\alpha _{2}=-\cos \phi \sin \psi +\,\sin \phi \,\sin \theta \cos \psi ,} \\
{\,\alpha _{3}=\,\sin \phi \sin \psi \,+\cos \phi \sin \theta \,\cos \psi ),}
\\
{\beta _{1}=\cos \theta \sin \psi ,} \\
{\,\,\beta _{2}=\cos \varphi \,\cos \psi \,+\sin \phi \sin \theta \sin \psi ,%
} \\
{\beta _{3}=\,-\sin \phi \cos \psi +\,\cos \phi \sin \theta \sin \psi ),} \\
{\gamma _{1}=-\sin \theta \,,} \\
{\gamma _{2}=\sin \phi \cos \theta \,,\,} \\
{\gamma _{3}=\cos \phi \cos \theta )},%
\end{array}
\label{GrindEQ__2_}
\end{equation}%
and
\begin{equation}
\mathbf{\vec{\alpha}}=\alpha _{1}i_{b}+\alpha _{2}j_{b}+\alpha _{3}k_{b}{,}%
\,\,\,\,{\vec{\beta}}=\beta _{1}i_{b}+\beta _{2}j_{b}+\beta _{3}k_{b}{,}%
\,\,\,\,{\vec{\gamma}}=\gamma _{1}i_{b}+\gamma _{2}j_{b}+\gamma _{3}k_{b},
\end{equation}
\begin{figure}[tbp]
\centering
\includegraphics[width=10cm, angle=0]{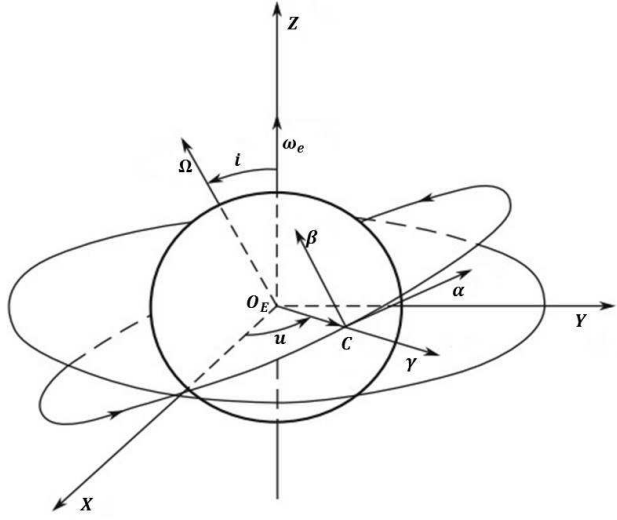}
\caption{Coordinates used in the derivation of the equations of motion }
\label{Fig1}
\end{figure}

\section{Torque due to  Lorentz Force}

The geomagnetic field with magnetic induction $\vec{B}$ is approximated by
the dipole approximation. The spacecraft is supposed to be equipped with a
charged surface (screen) of area $S$, with the electric charge $q=\int_{{S}%
}\sigma \,dS$ distributed over the surface with density $\sigma $.
Therefore, we can write the torque of these forces relative to the
spacecraft's center of mass as follows (Griffith \cite{Griffith})

\begin{equation}
{\vec{T}}_{{L}}=\int_{S}{\sigma }{\vec{\rho}}\times (\vec{V}\times {\vec{B})}%
{dS}.  \label{GrindEQ__3_}
\end{equation}%
where ${\vec{\rho}}$ is the radius vector of the screen's element ${dS%
}$ relative to the spacecraft's center of mass and $\vec{V}$ is the velocity
of the element ${dS}$ relative to the geomagnetic field. As in Tikhonov et.
al. (\cite{Tikhonov}), the torque ${\vec{T}}_{L}$ can be written as follows

\begin{equation}
{\vec{T}}_{L}=(T_{Lx},T_{Ly},T_{Lz})=q{\vec{\rho}}_{{0}}\times {A}^{T}({\vec{%
V}}_{{rel}}\times {\vec{B}}_{{o}}{)},  \label{GrindEQ__4_}
\end{equation}%
\begin{equation}
{\vec{\rho}}_{{0}}{=x}_{{0}}{i}_{{b}}{+y}_{{0}}{j}_{{b}}{+z}_{{0}}k_{{b}}{=}%
q^{-1}\int_{{S}}\sigma \,{\vec{\rho}}\,dS  \label{GrindEQ__5_}
\end{equation}%
${\vec{\rho}}_{{0}}$ is the radius vector of the charged center of a
spacecraft relative to its center of mass and ${A}^{{T}}$
 {is the transpose
of the matrix of the unitary vectors} $A$. As in Gangested (\cite%
{gangsted2010}), we use

\begin{equation}
{\vec{V}}_{{rel}}=(V_{rel1},V_{rel2},V_{rel3})={\vec{V}}-{\vec{\omega}}_{{e}%
}\times {\vec{r}}=\vec{r}{(}\Omega {-}\omega _{{E}}\,{\cos }i{)}\times {\vec{%
\alpha}+}R{\omega }_{{E}}\,{\sin i}\,{\cos u}\,{\vec{\beta}\;,\;}
\label{GrindEQ__6_}
\end{equation}%
where ${\vec{V}}_{{rel}}$ is the velocity vector of the spacecraft's center
of mass relative to the geomagnetic field, $\vec{V}$ is the initial velocity
of the satellite, ${\vec{\omega}}_{{e}}{=}\omega _{{e}}{\hat{\kappa}}$ is
the angular velocity of the diurnal rotation of the geomagnetic field
together with the Earth, ${\vec{B}}_{{o}}$ is the magnetic field in the
orbital coordinates.  { Substituting from equations (\ref{GrindEQ__3_}-\ref%
{GrindEQ__5_}) into equation (\ref{GrindEQ__6_}), }we can write the final
form of the components of the torque due to Lorentz force as below.
\begin{equation}
{{T}_{Lx}=q}\left\{
\begin{array}{c}
{y_{0}[\alpha _{3}V_{rel2}B_{o3}-\beta _{3}V_{rel1}B_{o3}+\gamma
_{3}(V_{rel1}B_{o2}-V_{rel2}B_{o1})]} \\
{-z_{0}(\alpha _{2}V_{rel2}B_{o3}-\beta _{2}V_{rel1}B_{o3}+\gamma
_{2}(V_{rel1}B_{o2}-V_{rel2}B_{o1})}%
\end{array}%
\right\} ,  \label{Tlx}
\end{equation}%
\begin{equation}
{{T}_{Ly}=q}\left\{
\begin{array}{c}
{z_{0}[\alpha _{1}V_{rel2}B_{o3}-\beta _{1}V_{rel1}B_{o3}+\gamma
_{1}(V_{rel1}B_{o2}-V_{rel2}B_{o1})]} \\
{-x_{0}(\alpha _{3}V_{rel2}B_{o3}-\beta _{3}V_{rel1}B_{o3}+\gamma
_{3}(V_{rel1}B_{o2}-V_{rel2}B_{o1})}%
\end{array}%
\right\} ,  \label{Tly}
\end{equation}%
\begin{equation}
{{T}_{Lz}=q}\left\{
\begin{array}{c}
{x_{0}[\alpha _{2}V_{rel2}B_{o3}-\beta _{2}V_{rel1}B_{o3}+\gamma
_{2}(V_{rel1}B_{o2}-V_{rel2}B_{o1})]\,} \\
{-y_{0}(\alpha _{1}V_{rel2}B_{o3}-\beta _{1}V_{rel1}B_{o3}+\gamma
_{1}(V_{rel1}B_{o2}-V_{rel2}B_{o1})}%
\end{array}%
\right\} .  \label{Tlz}
\end{equation}

As in Wertz (\cite{wertz1978}) we can write the components of the magnetic
field in the orbital system directed to the tangent of the orbital plane,
normal to the orbit, and in the direction of the radius respectively as
below.
\begin{eqnarray}
B_{o1} &=&\frac{{B}_{0}}{2r^{3}}\sin \theta _{m}^{\prime }\,[3\cos (2\nu
-\alpha _{m})+\cos \alpha _{m}],  \notag \\
B_{o2} &=&{-}\frac{{B}_{0}}{2r^{3}}\,\cos \theta _{m}^{\prime }\,,
\label{GrindEQ__8_} \\
B_{o3} &=&\frac{{B}_{0}}{2r^{3}}\sin \theta _{m}^{\prime }\,[3\sin (2\nu
-\alpha _{m})+\sin \alpha _{m}],  \notag
\end{eqnarray}%
where, ${B}_{0}=7.943\times 10^{15}$ is the intensity of the magnetic field,
$\theta _{m}^{\prime }=168.6^{\circ }$ is co-elevation of the dipole, and $%
\alpha _{m}=109.3^{\circ }$ is the east longitude of the dipole and $\nu $
is the true anomaly measured from ascending node.

\section{ Equilibrium positions and analytical Control Law}

The equations of motion of a rigid artificial satellite are usually written
in the Euler - Poisson variables ${\vec{\omega} }$, $\vec{\alpha }$,$\vec{%
\beta} $,$\vec{\gamma} $ and have the following form ( Abide-Ariz, \cite%
{yehia2007}).

\begin{equation}
\frac{d~{\vec{\omega}}}{dt}I+{\vec{\omega}}\times {\vec{\omega}}I={\vec{T}}%
_{G}+{\vec{T}}_{L},  \label{GrindEQ__9_}
\end{equation}%
\begin{equation}
\frac{d~\vec{\alpha}}{dt}\mathbf{+}\vec{\alpha}\mathbf{\times }\vec{\omega}%
\mathbf{=-}{\Omega \vec{\gamma},}\,\frac{d~{\vec{\beta}}}{dt}{+\vec{\beta}%
\times {\vec{\omega}}=0,}\,\,\frac{d~{\vec{\gamma}}}{dt}{+\vec{\gamma}\times
{\vec{\omega}}=\Omega \vec{\alpha}}  \label{GrindEQ__10_}
\end{equation}%
where, ${\vec{T}}_{G}=3\Omega ^{2}{\vec{\gamma}}\times {\vec{\gamma}}I$ is
well known formula of the gravity gradient torque. $I$ is the inertia matrix
of the spacecraft, $\Omega $ is the orbital angular velocity, $\vec{\omega}$
 is the angular velocity vector of the spacecraft. The components of $%
{\vec{T}}_{G}$ can be written as

 {\begin{equation}
\begin{array}{l}
{T}_{Gx}{=3\Omega ^{2}\gamma _{2}\gamma _{3}(C-B}){,} \\
{T}_{Gy}{=3\Omega ^{2}\gamma _{1}\gamma _{3}(A-C}){,} \\
{T}_{Gz}{=3\Omega ^{2}\gamma _{1}\gamma _{2}(B-A}){,}%
\end{array}
\label{TG}
\end{equation}
}
According to Gerlach (\cite{gerlach1965}), the angular velocity of the
spacecraft in the inertial reference frame is ${\vec{\omega}=\;(\omega }_{{x}%
}{,\omega }_{{y}}{,\omega }_{{z}})$, and in the orbital reference frame is ${%
\vec{\omega}}_{o}{=\;(\omega }_{ox}{,\omega }_{oy}{,\omega }_{oz}{)}$ where given below.

\begin{equation}
\begin{array}{l}
{{\omega }_{{x}}=\dot{\phi}-\dot{\psi}\sin \theta ,} \\
{{\omega }_{{y}}=\dot{\theta}\,\cos \phi +\dot{\psi}\cos \theta \,\sin \phi ,%
} \\
{{\omega }_{{z}}=-\dot{\theta}\,\sin \phi +\dot{\psi}\cos \theta \,\cos \phi
,}%
\end{array}
\label{GrindEQ__11_}
\end{equation}%
and%
\begin{equation}
\begin{array}{l}
{{\omega }_{{ox}}=\dot{\phi}-\dot{\psi}\sin \theta -\Omega \,\sin \psi \cos
\theta ,} \\
{{\omega }_{{oy}}=\dot{\theta}\,\cos \phi +\dot{\psi}\cos \theta \,\sin \phi
-\Omega (\cos \varphi \,\cos \psi \,+\sin \phi \sin \theta \sin \psi ),} \\
{{\omega }_{{oz}}=-\dot{\theta}\,\sin \phi +\dot{\psi}\cos \theta \,\cos
\phi -\Omega (-\sin \phi \cos \psi +\,\cos \phi \sin \theta \sin \psi )).}%
\end{array}
\label{GrindEQ__12_}
\end{equation}

It is well known that the orbital system rotate in space with a fixed orbital
angular velocity $\Omega $ about the axis, which is perpendicular to the
orbital plane. The relation between the angular velocity in the two systems
is $\vec{\omega }{=\vec{\omega} }_{{o}}-\Omega {\vec{\beta} .}$

 {At equilibrium positions, the right hand side of Eq.(\ref{GrindEQ__9_}) will
be zero. Substituting from Eqs.(\ref{Tlx}-\ref{Tlz}) and Eqs. (\ref{TG}) in
equation (\ref{GrindEQ__9_}) and after some algebraic manipulation we get
the following equilibrium positions.
}
\begin{itemize}
\item[Equilibrium 1.]
\begin{equation}
\theta =0,\phi =0,\psi =\frac{\pi }{{\ 2}},\left( \alpha _{1},\alpha
_{2},\alpha _{3}\right) =(0,-1,0),\,\,\,\,\left( \beta _{1},\beta _{2},\beta
_{3}\right) =(1,0,0),
\end{equation}%
\begin{equation}
\left( \gamma _{1},\gamma _{2},\gamma _{3}\right) =(0,0,1),
\end{equation}%
\begin{equation}
x_{0}=\frac{-V_{rel1}B_{o3}}{V_{rel1}B_{o2}-V_{rel2}B_{o1}}z_{0},\,\,\,y_{0}=%
\frac{-V_{rel2}B_{o3}}{V_{rel1}B_{o2}-V_{rel2}B_{o1}}z_{0}.
\label{equilibrium 1}
\end{equation}

\item[ Equilibrium 2.]
\begin{equation}
\theta =0,\phi =\frac{\pi }{{\ 2}},\psi =0,\left( \alpha _{1},\alpha
_{2},\alpha _{3}\right) =(1,0,0),\left( \beta _{1},\beta _{2},\beta
_{3}\right) =(0,0,-1),
\end{equation}%
\begin{equation}
\left( \gamma _{1},\gamma _{2},\gamma _{3}\right) =(0,0,1),
\end{equation}%
\begin{equation}
x_{0}=\frac{V_{rel2}}{V_{rel1}}z_{0},\,\,\,y_{0}=\frac{%
V_{rel1}B_{o2}-V_{rel2}B_{o1}}{V_{rel1}B_{o3}}z_{0}.  \label{equilibrium 2}
\end{equation}

\item[Equilibrium 3.]
\begin{equation}
\theta =\frac{\pi }{{\ 2}},\phi =0,\psi =0,\left( \alpha _{1},\alpha
_{2},\alpha _{3}\right) =(0,0,1),\left( \beta _{1},\beta _{2},\beta
_{3}\right) =(0,1,0),
\end{equation}%
\begin{equation}
\left( \gamma _{1},\gamma _{2},\gamma _{3}\right) =(-1,0,0),
\end{equation}%
\begin{equation}
x_{0}=\frac{V_{rel2}B_{o1}-V_{rel1}B_{o2}}{V_{rel2}B_{o3}}%
z_{0},\,\,\,\,\,y_{0}=\frac{-V_{rel1}}{V_{rel2}}z_{0}.  \label{equilibrium 3}
\end{equation}

\item[Equilibrium 4.]
\begin{equation}
\theta =0,\phi =0,\psi =0,\left( \alpha _{1},\alpha _{2},\alpha _{3}\right)
=(1,0,0),\left( \beta _{1},\beta _{2},\beta _{3}\right) =(0,1,0),
\end{equation}%
\begin{equation}
\left( \gamma _{1},\gamma _{2},\gamma _{3}\right) =(0,0,1),
\end{equation}%
\begin{equation}
x_{0}=\frac{V_{rel1}B_{o3}}{V_{rel1}B_{o2}-V_{rel2}B_{o1}}z_{0},\,\,\,y_{0}=%
\frac{V_{rel2}B_{o3}}{V_{rel1}B_{o2}-V_{rel2}B_{o1}}z_{0}.
\label{equilibrium 4}
\end{equation}
\end{itemize}

It can be seen that the four equilibrium positions depend on $z_{0} $ which
can control the equilibrium positions. We will study the relationship
between the magnitude of the torque, magnitude of the radius vector of the
charged center of spacecraft relative to its center of mass, the amount of
charge, and the inclination of the orbits. This analysis will be done for
two different values of $z_{0}$,

\begin{itemize}
\item $z_{0}=k\,B_{o2},\,\,\,k={-}\frac{2r^{3}}{{B}_{0}}$ which
approximately equal unity (1 meter)

\item $z_{0} =4$
\end{itemize}

\section{Numerical results}

\subsection{Equilibrium 1}

In this equilibrium position the attitude motion of satellite is in the $%
\psi $ direction only. The magnitude of the radius vector $\vec{\rho}_{0}$
is given by $\left\Vert \vec{\rho}_{0}\right\Vert =\sqrt{%
x_{0}^{2}+y_{0}^{2}+z_{0}^{2}}$. In case of equilibrium 1, the values of $%
x_{0},$ and $y_{0}$ can be determined from equation (\ref{equilibrium 1})
which will give the magnitude of $\vec{\rho}_{0}$ as a function of ${u,i,}$
and ${z_{0}.}$
\begin{equation}
{\rho }_{0}{(u,i,z_{0})=}\left\Vert \vec{\rho}_{0}\right\Vert =z_{0}\sqrt{%
\begin{array}{c}
1{+}2.98{\times }{10}^{30}{\left( \frac{{-1.1}{{\times }10}^{-3}{+}7.27{{%
\times }10}^{-5}{\cos }\left( i\right) }{{D}_{{eq1}}}\right) }^{2} \\
{+}1.57{\times }{10}^{22}{\left( \frac{{\cos }\left( u\right) {\sin }(i)}{{D}%
_{{eq1}}}\right) }^{{2}}%
\end{array}%
},  \label{rhoeq1}
\end{equation}%
where%
\begin{equation}
{D}_{{eq1}}=2.07{\times }{10}^{12}{-}1.37{\times }{10}^{11}{\cos (}i){+}2.83{%
\times }{10}^{11}{\cos (}u){\sin (}i).
\end{equation}%
Similarly the magnitude of torque ${\vec{T}}_{L}$ can be determined from
equations (\ref{GrindEQ__4_}) to (\ref{Tlz}).
\begin{equation}
\left\Vert {\vec{T}}_{L}{(q,u,i},{r)}\right\Vert {=}\frac{{qz}_{0}}{r^{{2}}{D%
}_{{eq1}}}\sqrt{%
\begin{array}{c}
{\cos }^{2}u{\sin }^{2}i(2.52\times 10^{15}+1.10\times 10^{13}{\cos }^{2}i
\\
+2.84\times 10^{14}{\cos }u{\ \sin }i+1.95\times 10^{13}{\cos }^{2}u{\ \sin }%
^{2}i \\
+{\ \cos }i(-3.33\times 10^{14}-1.88\times 10^{13}{\cos }u{\ \sin }i))%
\end{array}%
},  \label{torqueEq1}
\end{equation}

\begin{figure}[tbp]
\centering
\resizebox{130mm}{!}{\includegraphics {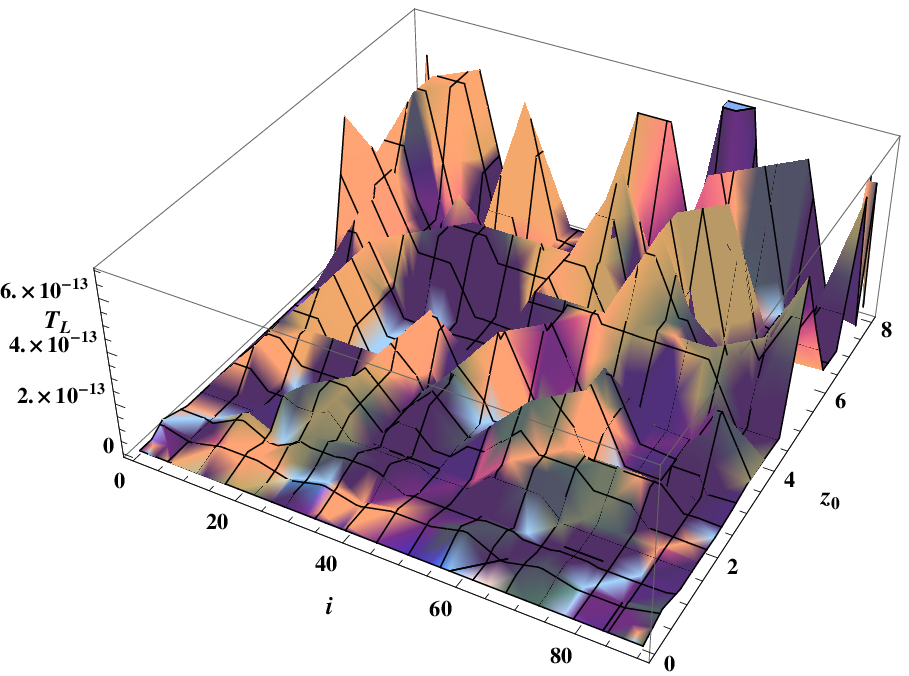}\includegraphics
{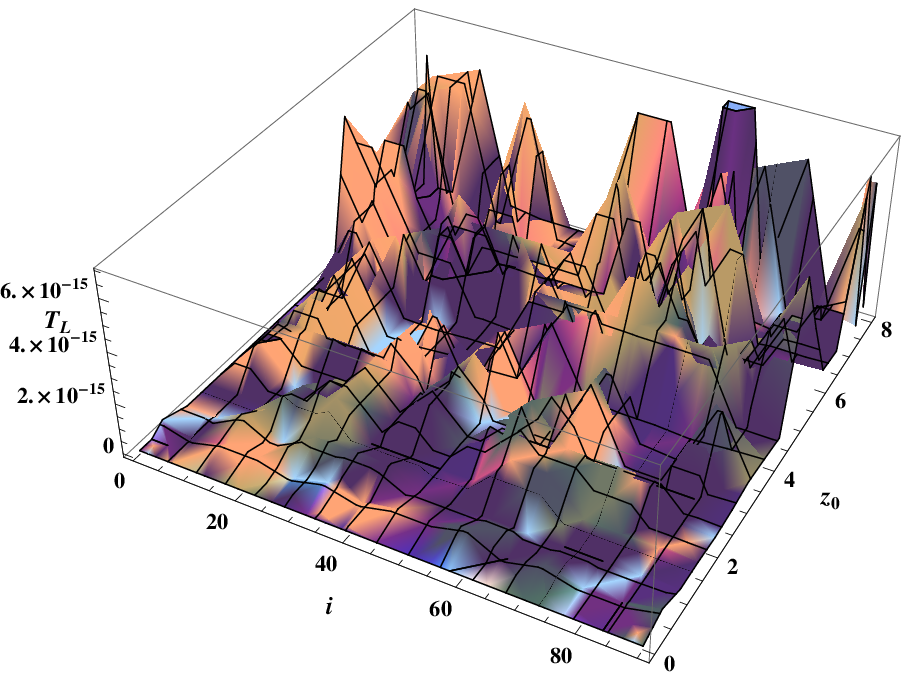}}
\caption{ $\left\Vert {\vec{T}}_{L}{(q,u,i,r)}\right\Vert $ for fixed values
of charge $q=10000C(left),100C (right)$, $r=6900km$ and $u=40$ in the case
of equilibrium 1}
\label{Torque-i-z0-eq1}
\end{figure}

It can be seen from equations (\ref{rhoeq1}) that $\Vert \vec{\rho}_{0}{%
(u,i,z_{0})}\Vert $ is independent of $r$ even though its components depend
on it. Equation (\ref{torqueEq1}) gives the magnitude of the torque. $\Vert
\vec{\rho}_{0}{(u,i.z_{0})}\Vert $ is an almost periodic function of
inclination $i$ and latitude $u$ with a maximum value of 1.4029 meters and
minimum value of 1.236 meters for $z_{0}=kB_{o2}=0.96$. As the function is
almost periodic therefore these optimum values occur at various values of $i$
and $u$. For example the maximum occurs at $(i,u)=$ $(23.63,21.99)$ and $%
(58.05,53.41)$. Similarly the minimum occurs at $(i,u)=(39.00,37.70)$, and $%
(58.05,56.55)$. To see the dependence of $\Vert \vec{\rho}_{0}{(u,i,z_{0})}%
\Vert $ on the inclination $i$ and latitude $u$, please refer to figure (\ref%
{equib1} ). It can be seen both from equation (\ref{rhoeq1}) and figure (\ref%
{equib1}) that $z_{0}$ can be used to control $\rho $. In a similar way $%
z_{0}$ can be used to control torque as can be seen in equation (\ref%
{torqueEq1}) . The relationship of Torque with $r$ and $q$ is
straightforward. It can be seen from equation (\ref{torqueEq1}) that the
torque is directly proportional to $q$ and inversely proportional to $r^{2}$%
. Figure (\ref{Torque-i-z0-eq1} ) also shows that $q$ can be used to control
the torque if desired. It can also be seen from figure (\ref{Torque-i-z0-eq1}
) which is given for fixed values of $q,u$ and $r$ that torque has a maximum
value of the order $10^{-13}$ for each value of inclination $i.$

\begin{figure}[tbp]
\centering
\resizebox{120mm}{!}{\includegraphics{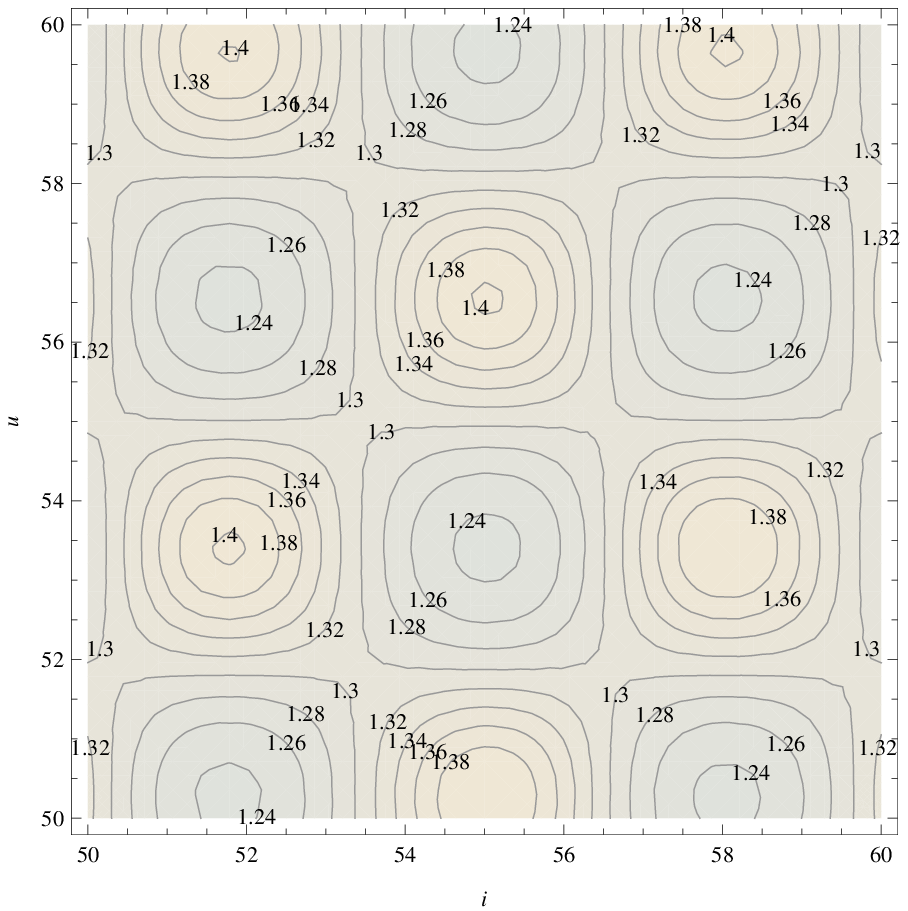}
\includegraphics{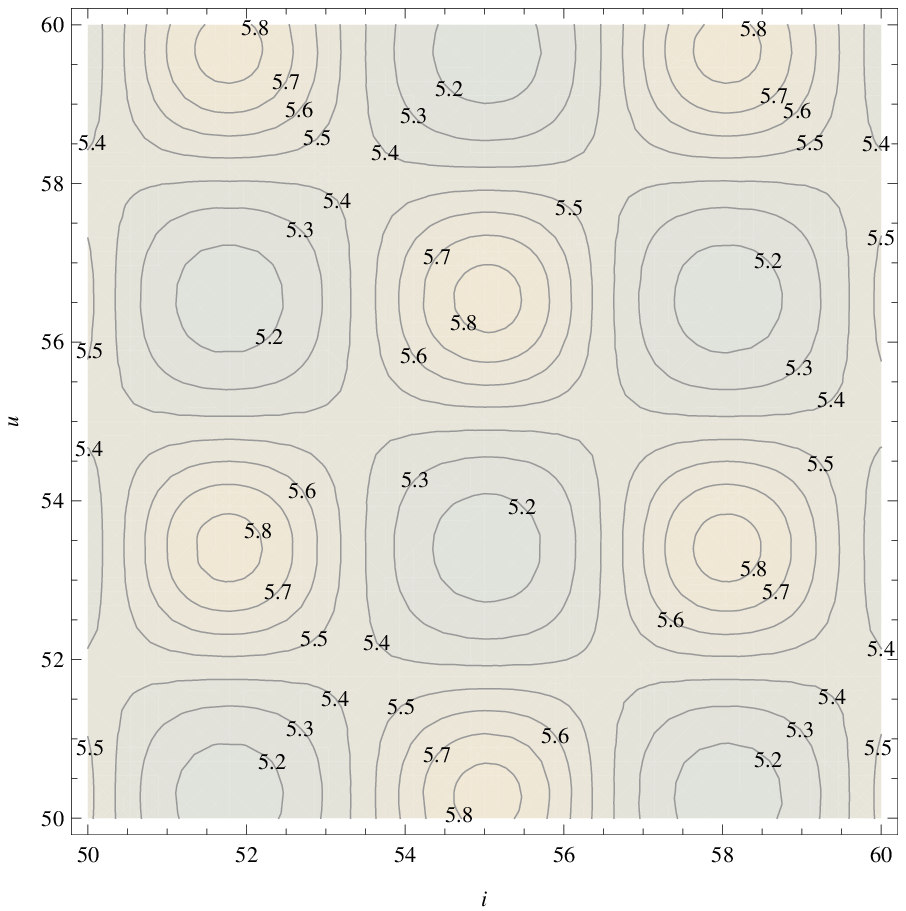}}
\caption{{a. Contour plot of $\|{\protect\rho }_0{(u,i,z_0)}\| $ with
maximum and minimum values occurring more than once confirming its periodic
behavior. $z_0$ is taken to be 0.96 in the case of equilibrium 1 b. Contour
plot of $\|{\protect\rho }_0{(u,i,z_0)}\| $ with maximum and minimum values
occurring more than once confirming its periodic behavior. $z_0$ is taken to
be 4 in the case of equilibrium 1.}}
\label{equib1}
\end{figure}

\subsection{Equilibrium 2}

\begin{figure}[tbp]
\centering
\resizebox{130mm}{!}{\includegraphics
{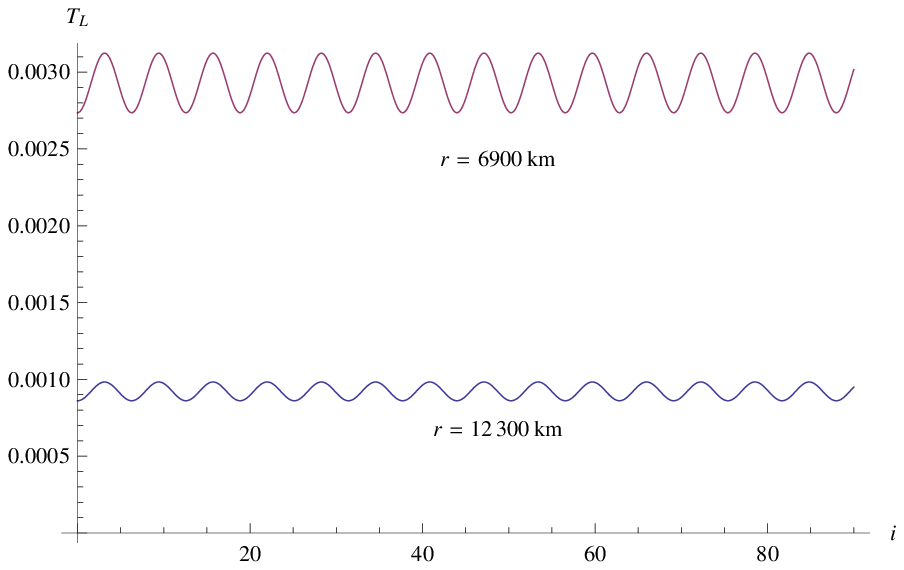}\includegraphics{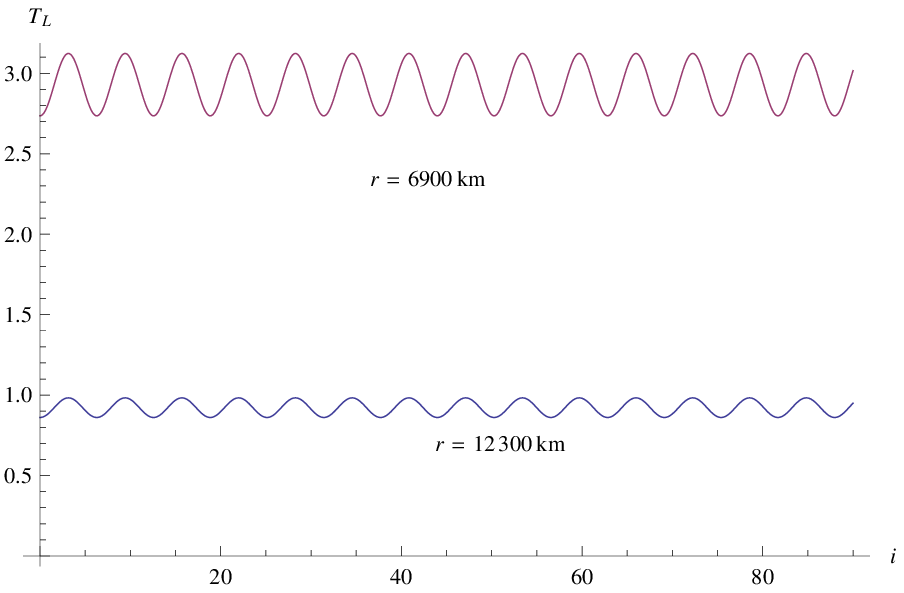}}
\caption{ $\left\Vert {\vec{T}}_{L}{(q,u,i,r)}\right\Vert $ for fixed value
of charge $q=0.01$ (left), $q=10$ (right) and $z_{0}=1$ in the case of
equilibrium 2 }
\label{eq2-1}
\end{figure}

In this equilibrium position the attitude motion of satellite is in the roll
direction only. In this case $\Vert \vec{\rho}_{0}(z_{0})\Vert $ is a linear
function of $z_{0}$ only. It has a value of $2.47z_{0}$. Torque is a
function of the inclination $i$, charge $q$ and $r$ only.
\begin{equation}
\left\Vert {\vec{T}}_{L}{(q,i,r)}\right\Vert =1.27\times 10^{16}\frac{z_{0}}{%
r^{2}}|q(0.0011-0.0000727\cos i)|
\end{equation}%
In the same way as in equilibrium one, it is directly proportional to $q$
and inversely proportional to $r^{2}$. Unlike equilibrium one, Torque in
this case is a periodic function of the inclination $i$ for fixed values of $%
q$ and $r^{2}$. For fixed values of charge $q=0.01C$,or $q=10C,$ $z_{0}=1$,
and $r=6900km$ or $r=12300$ the optimum values of torque changes
periodically. To see the periodic behavior of the torque and a comparison of
the torque for two different values of $r$, see figure (\ref{eq2-1}). From
the comparison for $r=6900km$ and $r=12300km$ we can see that the value of
the Lorentz torque is higher in Low Earth Orbits (LEO). When charge is
increased from $0.01C$ to $10C$ the magnitude of Lorentz torque increase
significantly. It means electrostatic charge can be used as some type of
control if desired. This can be seen in figure (\ref{eq2-1}).

\subsection{Equilibrium 3}

In this case $\Vert \vec{\rho}_{0}\Vert $ is a linear function of $z_{0}$.
It has a value of $1.42138z_{0}$. Torque in this case is zero. The attitude
motion of the satellite is in the pitch direction and the electrostatic of
the screen surface is almost constant which makes the components of Lorentz
Torque zero.

\subsection{Equilibrium 4}

This position is a special case which can happen only when the orbital
system coincides with the principal axis of inertia which is rigidly fixed
to the satellite. For equilibrium 4 described in section 4, $\Vert \vec{\rho}%
_{0}\Vert $ and $\Vert {\vec{T}}_{L}\Vert $ {are} determined in the same way
as in the case of equilibrium 1.

\begin{eqnarray}
\Vert \vec{\rho}_{0}{(u,i,z_{0})}\Vert &=&z_{0}\sqrt{1+\frac{%
2.98(0.11-0.727\cos i)^{2}}{D_{eq4}}+\frac{1.57(\cos u\sin i)^{2}}{D_{eq4}}},
\label{rhoEq4} \\
D_{eq4} &=&(42.8-2.83\cos i-1.37\cos u\sin i)^{2}.  \notag
\end{eqnarray}%
%
%
%
%
%
\begin{equation}
\left\Vert {\vec{T}}_{L}{(q,u,i,z}_{0},{r)}\right\Vert {=}\frac{{qz_{0}}%
\times 10^{11}}{r^{{2}}}\sqrt{%
\begin{array}{c}
(19-1.25(\cos i+\cos u\sin i))^{2} \\
+(19-1.25(\cos i-\cos u\sin i))^{2} \\
+\left( 3.6-47.6\cos i+1.57\cos ^{2}i+1.57\cos ^{2}u\sin ^{2}i\right) ^{2}%
\frac{1}{D_{eq4}}%
\end{array}%
}.  \label{torqueEq4}
\end{equation}

\begin{figure}[tbp]
\centering
\resizebox{140mm}{!}{\includegraphics{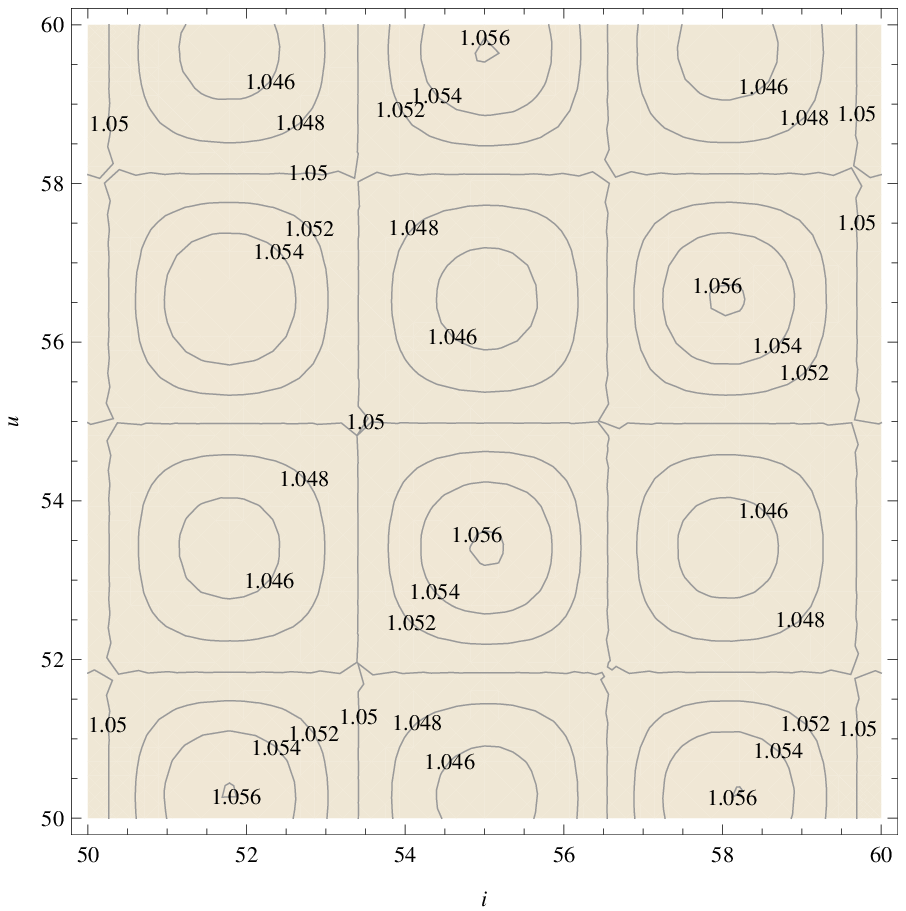}
\includegraphics{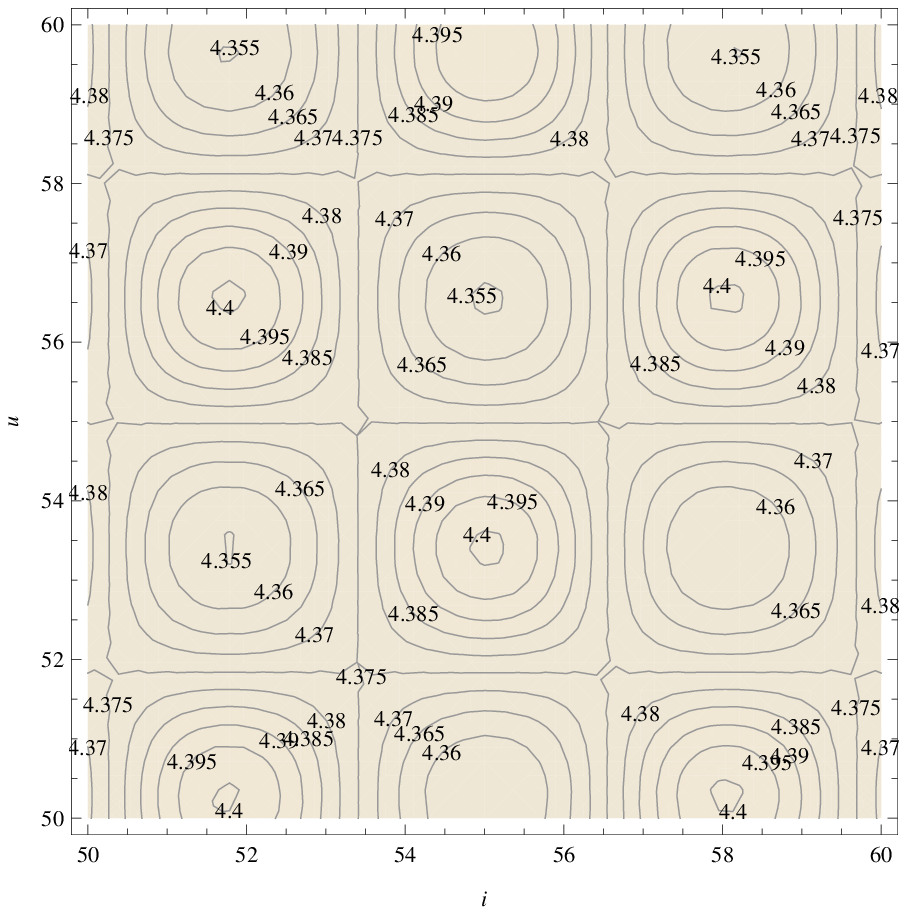}}
\caption{{a. Contour plot of $\|{\protect\rho }_0{(u,i)}\| $ with maximum
and minimum values occurring more than once confirming its periodic
behavior. $z_0$ is taken to be 0.96 ( equilibrium 4) b. Contour plot of $\|{%
\protect\rho }_0{(u,i)}\| $ with maximum and minimum values occurring more
than once confirming its periodic behavior. $z_0$ is taken to be 4 (
equilibrium 4).}}
\label{fig-rho-eq4}
\end{figure}

\begin{figure}[tbp]
\centering
\resizebox{140mm}{!}{\includegraphics {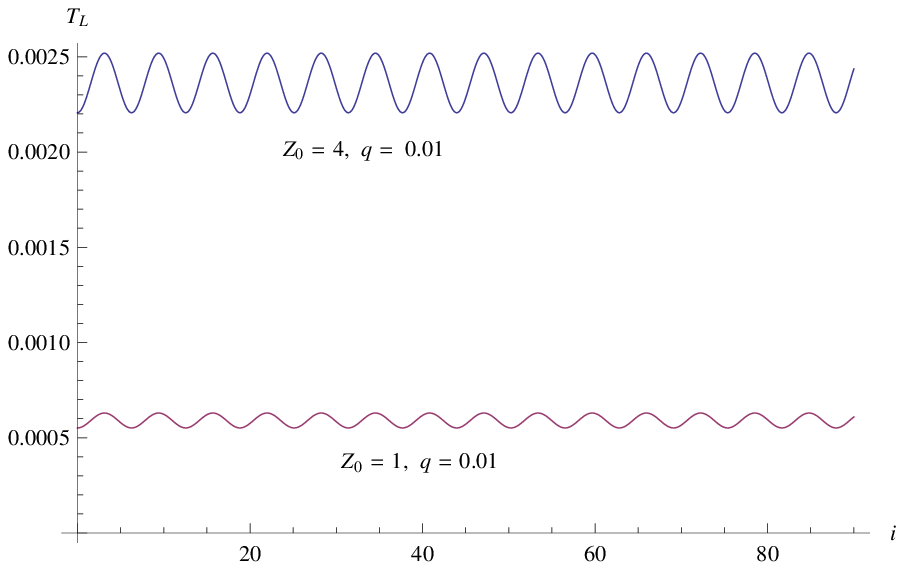}\includegraphics
{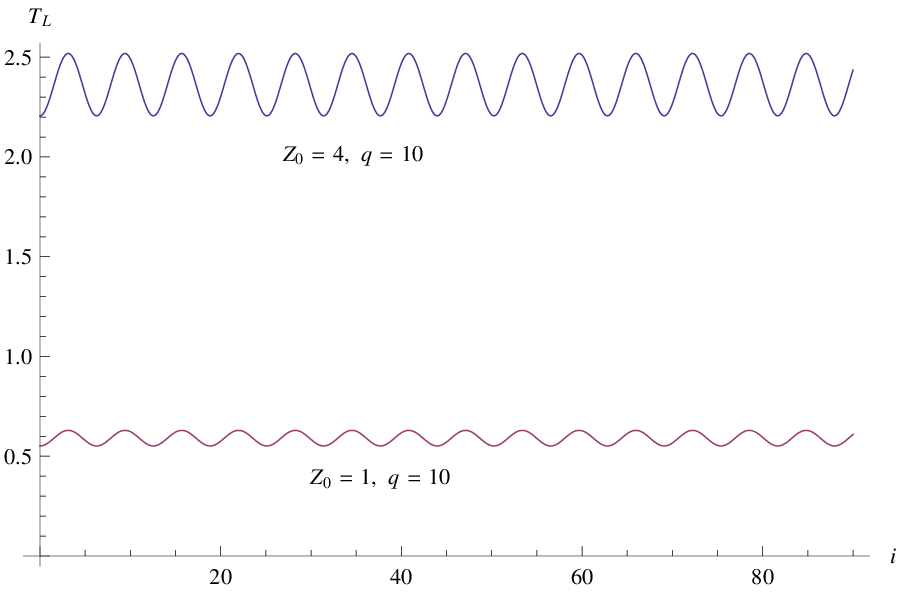}}
\caption{ $\left\Vert {\vec{T}}_{L}{(q,u,i,r)}\right\Vert $ for fixed value
of altitude ($r=6900km$, latitude ($u=20$) and two different values of $%
z_0=1,z_0=4$,and $q=0.01C,q= 10C$ in the case of equilibrium 4 }
\label{torq-eq4-1}
\end{figure}

It can be seen from equations (\ref{rhoEq4}) that $\Vert \vec{\rho}_{0}{%
(u,i,z_{0})}\Vert $ is independent of $r$ even though its components depend
on it. Equation (\ref{torqueEq4}) gives the magnitude of torque. $\Vert \vec{%
\rho}_{0}{(u,i,z_{0})}\Vert $ is an almost periodic function of inclination $%
i$ and latitude $u$ with a maximum value of 1.057 meters and minimum value
of 1.04611 meters for $z_{0}=0.96$. As the function is almost periodic
therefore these optimum values occur at various values of $i$ and $u$. For
example the maximum occurs at $(i,u)=(61.33,34.56)$ . Similarly the minimum
occurs at $(i,u)=(58.05,53.40)$. For some other occurrences of the optimum
values, see figures (\ref{fig-rho-eq4}). It can be seen both from equation (%
\ref{rhoEq4}) and figure (\ref{fig-rho-eq4}) that $z_{0}$ can be used to
control $\rho _{0}$. In a similar way $z_{0}$ can be used to control torque
as can be seen in equation (\ref{torqueEq4}). The relationship of Torque
with $r$ and $q$ is straightforward. It can be seen from equation (\ref%
{torqueEq4}) that the torque for equilibrium four is directly proportional
to $q$ and inversely proportional to $r^{2}$. Therefore $q$ and $z_{0}$, can
be used to control torque if desired. To completely describe the torque, its
representative graph is given in figures (\ref{torq-eq4-1} ). In the same
way as in equilibrium 2 when charge is increased from $0.01C$ to $10C$ the
magnitude of Lorentz torque increases significantly. It means electrostatic
charge can be used as some type of control if desired which can be seen in
figures (\ref{torq-eq4-1} ). It can also be seen from figure (\ref%
{torq-eq4-2} ) which is given for fixed values of $q=0.01C,z_{0}=2$ and $%
r=6900km$ that torque has a maximum value of the order $10^{-2}$ for each
value of inclination $i.$

\begin{figure}[tbp]
\centering
\resizebox{100mm}{!}{\includegraphics
{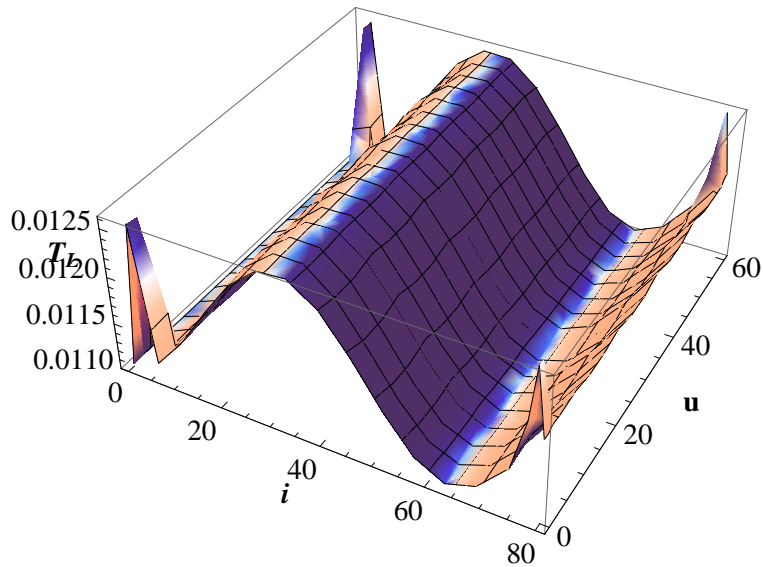}}
\caption{ $\left\Vert {\vec{T}}_{L}{(q,u,i,r)}\right\Vert $ for fixed values
of $q=0.01$, $z_{0}=2$ and $r=6900km$ in the case of equilibrium 4 }
\label{torq-eq4-2}
\end{figure}

\section{Conclusions}

To control the attitude of a general shape charged satellite we proposed the
utilization of a Lorentz torque with the gravity gradient torque . The
effect of Lorentz torque on the attitude dynamics and the orientation of the
equilibrium positions is discussed. The satellite is assumed to move in a
circular orbit in the geomagnetic field. For this particular setup we
derived four equilibrium positions. The attitude motion for these
equilibrium positions is analyzed in detail for different values of charge ($%
q)$, charged center of the satellite relative to its center of mass $(\rho
_{0})$, inclination, and latitude. The numerical results confirm that the
Lorentz torque has a significant effect on the attitude orientation of
satellite for any inclination, specially in highly inclined orbits.

In the case of equilibrium 1, 2 and 4, it is shown that the value of charge $%
q$ can control the magnitude of the Lorentz torque. We can choose the
optimal torque to create natural force which can be used to control the
attitude of the satellite. In case of equilibrium 1, a very high amount of
charge is needed to generate a reasonable amount of torque. That is, a $1000C
$ charge is needed to generate Lorentz torque of the order $10^{-13}.$ On the
other hand, in case of equilibrium 2 and 4 a charge of $0.01C$ will generate
a torque of the order $10^{-3}.$ This means that the use of charge as a control
is a more realistic option in equilibrium 2 and equilibrium 4. This also
means that, Lorentz force can be used to control satellite without consuming
too much propellant. The installation of such control on a satellite is
dependent on the size of the surfaces of the satellite, and the screen
charging, which can be realized by manufacturing a system of electrodes
simulating the controlled electrostatic layer. Such kind of control may be
used instead of the magnetic control system, as it is easy to control
the mass of the satellite and decrease the cost.

\end{document}